\documentclass[usenatbib]{mn2e}
\usepackage{epsfig}
\usepackage{amsmath,amssymb}
\newcommand {\rs} {$R_{\rm s}$}

\newcommand {\rvir} {$R_{\rm vir}$}

\newcommand{\hMsol}{{\,\rm h^{-1}M}_\odot}

\newcommand{\hMpc}{{\,\rm h^{-1}Mpc}}
\newcommand{\hkpc}{{\,\rm h^{-1}kpc}}
\newcommand{\G}{{\rm G}}

\title{Entropy of gas and dark matter in galaxy clusters} 
\author[Faltenbacher et al.]
{\parbox[t]\textwidth{Andreas Faltenbacher$^{1,2}$, Yehuda Hoffman$^3$, 
Stefan Gottl\"ober$^4$ and Gustavo Yepes$^5$}
\vspace*{6pt} \\
$^1$ Shanghai Astronomical Observatory,
80 Nandan Road, Shanghai 200030, 
China\\
$^2$UCO/Lick Observatory,
University of California at Santa Cruz, 
1156 High Street, Santa Cruz, CA 95064, USA
\\
$^3$Racah Institute of Physics, 
Hebrew University, 
Jerusalem 91904, Israel 
\\
$^4$Astrophysikalisches Institut Potsdam,
An der Sternwarte 16, 14482 Potsdam, Germany
\\
$^5$Grupo de Astrof\'{\i}sica, 
Universidad Aut\'onoma de Madrid,
Madrid E-280049, Spain 
\\
}
\date{\today}
\begin{document}
\maketitle
\begin{abstract}
On the basis of a large scale 'adiabatic', namely  non-radiative and
non-dissipative, cosmological smooth particle hydrodynamic simulation
we compare the entropy profiles of the gas and the dark matter (DM)
in galaxy clusters. We employ the quantity $K_{\rm g}= 3k_BT_{\rm
g}\rho{_{\rm g}^{-2/3}}/(\mu m_p) = \sigma^2_{\rm g}\rho{_{\rm
g}^{-2/3}}$ as measure for the {\it entropy} of the  intra-cluster
gas. By analogy the DM entropy is defined as $K_{\rm DM} =
\sigma{^2_{\rm DM}}\rho{_{\rm DM}^{-2/3}}$ ($\sigma{^2_{\rm DM}}$
is the 3D velocity dispersion of the DM). The DM
entropy is related to the DM phase space density by $K_{\rm DM}\propto
Q{_{\rm DM}^{-2/3}}$. In accord with other studies the radial DM phase
space density profile follows a power law behaviour, $Q_{\rm DM}
\propto r^{-1.82}$, which corresponds to $K_{\rm DM}\propto
r^{1.21}$. The simulated intra-cluster gas  has a flat entropy core
within $(0.8 \pm 0.4) R_{\rm s}$, where \rs\ is the NFW scale radius.
The outer profile follows the DM behaviour, $K_{\rm g}\propto
r^{1.21}$, in close agreement with X-ray observations. Upon scaling
the DM and gas densities by their mean cosmological values we find
that outside the entropy core a constant ratio of $K_{\rm g}/K_{\rm
DM}=0.71\pm0.18$ prevails. By extending the definition of the gas
temperature to include also the bulk kinetic energy the ratio of the
DM and gas extended entropy is  found to be unity for $r\gtrsim 0.8
R_{\rm s}$. The constant ratio of the gas thermal entropy to that of
the DM implies that observations of the intra-cluster gas can provide
an almost direct probe of the DM.
\end{abstract}
\begin{keywords}
methods: numerical - galaxies: clusters: general
\end{keywords}
\section{Introduction}
Clusters of galaxies represent a composite of baryonic and dark
matter (DM). The gas is commonly described by its thermodynamic
variables, such as density, temperature, pressure and also
entropy. The entropy provides a unique gauge for monitoring the
thermal history of the gas. In the standard model of structure
formation the primordial gas is isentropic. Ignoring radiative and
dissipative processes and other non-gravitational sources or sinks of
energy the only mechanism by which the gas entropy can change is {\it
via} shock waves. According to the
paradigm of hierarchical structure formation clusters form by 
merging and accretion of smaller building blocks, which usually
contain baryons and DM. The collapse and merging of
substructures lead to emergence of shock waves in the forming
intra-cluster gas (ICM) and it ceases to be isentropic. These shocks
can be seen close to the virial radius when the gas is accreted
onto the cluster (see e.g. ~\citealt{Borgani05, Voit05, 
Voit05b}). However, the study of \cite{Pfrommer06} shows that 
a non negligible amount of energy is dissipated by shocks further
inside the cluster potential well. In any case the resulting entropy
structure reflects the contribution of irreversible processes in the
formation of clusters. The main problem addressed here is to what
extent the gas entropy resembles the structure of the DM halo and
whether the classical thermodynamic description of the gas can be
extended and applied to the DM. 
The entropy of the ICM can be deduced from X-ray observations. 
As has become customary in this field, we refer to $K\propto
T_{\rm g}\rho_{\rm g}^{-2/3}$ as the 'entropy', where $T_{\rm g}$ is
temperature and $\rho_{\rm g}$ is the density of the gas.
More precisely, we define $K_{\rm g}\equiv 3k_BT_{\rm
g}\rho{_{\rm g}^{-2/3}}/(\mu m_p) = \sigma^2_{\rm g}\rho{_{\rm
g}^{-2/3}}$, where $k_B$ is the Boltzmann constant and $\sigma_{\rm
g}$ is the internal 3D velocity dispersion of the gas. It is the
latter expression upon which the direct comparison between
the gas and the dark matter is based.
Using {\it XMM-Newton} observations of 10 relaxed clusters
\cite{Pratt06} find the radial profile of $K$ follows a power law of
$r^\beta$ with $\beta \approx 1.1$ (see also \citealt{Donahue06}). 
Apart from the central regions the
slope obtained in adiabatic simulations agree with the observed values
(e.g.~\citealt{Voit05}). The observations, however, show an excess in
entropy which is thought to originate from non gravitational processes
(e.g. \citealt{Finoguenov03, Borgani05}). Note, the actual entropy of
an ideal gas is $S=\ln(K^{3/2})+ const.$, but here we follow the
common convention and associate the entropy with $K$.  
The concept of entropy has also been applied to
collisionless self-gravitating systems  in its Boltzmann-Gibbs
formulation (e.g.~\citealt{LyndenBell67,LyndenBell68})
and in an analogy with the thermodynamic entropy of a perfect gas
\citep{White87,Navarro95,Eke98}. It is the latter approach that we
follow here. In  analogy with classical thermodynamics and the ideal
gas equation of state we write  the local specific entropy 
$S_{\rm DM}=\ln(\sigma{_{\rm DM}^3} / \rho_{\rm DM})$,
where $\sigma_{\rm DM}$ is the DM velocity dispersion and an additive constant is  
neglected. A similar approach is presented by \cite{Eke98}.  Following
the convention used in the studies of clusters of galaxies we define
$K_{\rm DM}=\sigma{^2_{\rm DM}}\rho{^{-2/3}_{\rm DM}}$. Note that the
DM specific entropy  is the log of the
inverse of the so-called phase space density
$Q_{\rm DM}=\rho_{\rm DM} \sigma_{\rm DM}^{-3}$. Now, the analysis of
simulated DM halos by \cite{Taylor01} has shown that the radial phase
space density profile closely follows a power law, $Q_{\rm
DM}(r)\propto r^{-1.875}$. Using a set of high-resolution N-body 
simulations \cite{Ascasibar04} confirm this power-law behaviour for an
extend range of haloe masses, in addition they find coincidence of the
zero-points if the profiles are conveniently rescaled by virial
quantities. Changing variables from $Q$ to $K$ the \cite{Taylor01}
power law is given by   
\begin{equation}
K_{\rm DM}(r)=Q{_{\rm DM}^{-2/3}}(r) \propto r^{1.25}.
\end{equation}
The close similarity of the (appropriately converted) \cite{Taylor01}
slope and the slope derived from X-ray observations by \cite{Pratt06}
has motivated us to compare the DM and gaseous entropy
profiles of clusters of galaxies extracted from a large scale
'adiabatic' Smoothed Particle Hydrodynamics (SPH) simulation. 
In this context adiabatic is defined as non-radiative with no
non-gravitational sources or sinks of energy. In such kind of
simulations the entropy of each particle remains constant as long as
it does not pass through a shock front \citep{Springel05}. The only
dissipative process is introduced by the numerical viscosity which
damps unphysical oscillations behind shock fronts and thus transfers
kinetic into thermal energy in accordance with the Rankine-Hugoniot
jump conditions.  
This paper is organised as follows. \S\ref{sec:sim} introduces the
simulation and the cluster ensemble and ~\S\ref{sec:profiles} compares
the entropy profiles of the gaseous and the DM component of the
clusters. \S \ref{sec:core} focuses on the emerging gas entropy cores
and discusses their physical properties.  And a discussion is
given in ~\S\ref{sec:discu}.      
\section{Simulation}
\label{sec:sim}
This simulation, dubbed {\sc The MareNostrum
Universe} and described in detail in \cite{Gottloeber06}, 
was performed with the entropy conserving {\sc
Gadget2} code \citep{Springel05}. It followed the non linear evolution
of the cosmic density fields of the gas and the DM from
$z=40$ to the present epoch ($z=0$) within a comoving cube of $500\hMpc$
edges. The spatially flat concordance model employing the following 
parameters was adopted: $\Omega_m=0.3$, $\Omega_\Lambda=0.7$,
$\Omega_b=0.045$, the Hubble parameter $h=0.7$, the slope of the power
spectrum $n=1$ and the normalisation $\sigma_8=0.9$. Both components,
the gas and the DM, are resolved by $1024^3$ particles, which results
in a mass of $8.3 \times 10^9\hMsol$ for the DM particles and $1.5
\times  10^9\hMsol$ for the gas particles. The gas dynamical equations
were solved by SPH without including
dissipative or radiative processes or star formation, i.e. the gas was
treated adiabatically in the above discussed sense. The SPH smoothing
length has been set to the $40^{th}$ nearest neighbour of each SPH
particle and a gravitational force resolution was chosen equivalent to
a Plummer softening of $15\hkpc$ (comoving). The SPH softening was not
allowed to fall below the gravitational softening length.  

The cluster ensemble comprises 96 massive clusters selected from the
simulation volume. The clusters were found by a friends of friends
algorithm (FoF) with a linking length of 0.17 times the mean particle
separation applied to the DM component only. This corresponds roughly
to the virial overdensity. In a second step substructures within the
cluster volumes were identified by reducing the linking length of the
FoF group finder by a factor of 4 which defines substructures at
roughly 64 times the virial overdensity. 
The choice of this particular
linking length is somewhat arbitrary, however it has been found
to adequately identify relevant substructure, such as double or
triple core systems.
The cluster centre was associated with the centre
of the most massive substructure of the cluster. Subsequently the virial radius
\rvir\  determined as the cluster centric radius which encloses 334
times the mean cosmic density and the virial mass were computed as the 
sum of the gas and the dark matter mass within that radius. 
The virial masses lie in between $3.8\times 10^{14}$ and $ 2.5\times
10^{15}\hMsol$ corresponding to $\sim 5\times10^4-2\times10^5$
particles per species. In total 370 clusters with masses above
$3.8\times 10^{14}\hMsol$ are found within the simulation volume and
for a even lower limit of $10^{14}\hMsol$ we obtain 3708 objects.  
The selection process for the clusters in combination with the steep
mass function causes our sample to thin out at the low mass end, but
for high masses ($\gtrsim 5\times 10^{14}\hMsol$) it can assumed to 
be complete. To assure the best possible determination of the central
properties of the cluster the centre was redefined as the associated
peak of the density field smoothed by a $100\hkpc$ sphere. On average
this re-adjustment changed the location of the centres by only a few
$\hkpc$. 
\section{Entropy profiles of the gas and the dark matter}
\label{sec:profiles}
\begin{figure*}
\epsfig{file=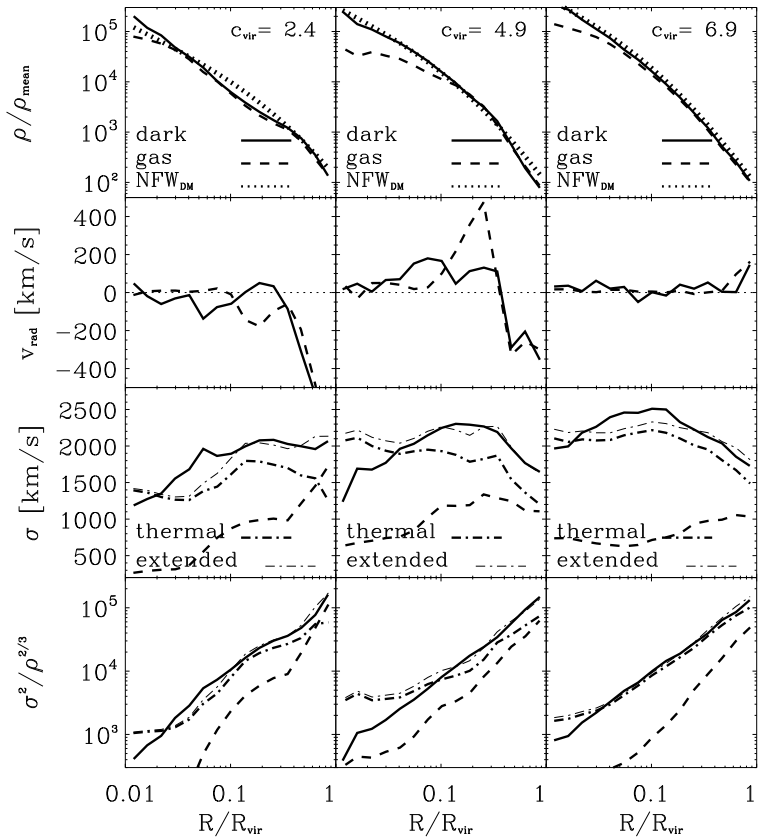,width=0.6\hsize}
\caption{\label{fig:select3}
Radial profiles for three arbitrarily selected clusters. The uppermost
panels display the density profiles of the DM (solid line) and
the gas (dashed line) scaled by their mean cosmic values. Here $c$
indicates the DM concentration derived from a NFW-fit (dotted
line). The second row shows the mean radial velocities for the dark
matter (solid line) and the gas (dashed line). The third row displays
the velocity dispersions $\sigma$. The DM dispersion is given
by the solid line. The turbulent dispersion of the gas is shown as
dashed line. The thick dashed-dotted line depicts the thermal velocity
dispersion of the gas and the thin dashed line displays the extended
one. The lowest panels show the DM and gas entropy. The same line
styles and units are used as in the panels above. 
}
\end{figure*}
The aim of this section is to demonstrate the similarities in the
entropy profiles of the dark and the gaseous matter. As motivated
above we will refer to the entropy of the gas and the DM as   
\begin{equation}
\label{equ:Ki}
K_i = {\sigma_i^2\over\rho_i^{2/3}}\ , 
\end{equation}
were $\rho$ is the density, $\sigma$ is the three dimensional
velocity dispersion and $i$ stands for either $DM$ (dark matter) or
$g$ (gas). The velocity dispersion of the dark matter within a radial
bin is computed as the rms velocity of the particles within
that bin. By their nature, the gas particles are characterised by two
distinct properties related to their kinetic energy, namely, a proper
velocity (same as for dark matter particles) and a temperature. The
proper motions of the gas particles will be referred to as bulk
motions and the corresponding velocity dispersion will be addressed as
{\it turbulent dispersion}. Also the temperature of the gas will be
expressed by means of the velocity dispersion that represents the
thermal energy per unit mass, namely by $\sigma^2 = 3 k_B T /(\mu
m_p)$, where $m_p$ is the proton mass and $\mu=0.588$ for a completely
ionised gas. The velocity dispersion of the gas derived from its
temperature will be referred to as {\it thermal dispersion}. The
composite of the two gaseous dispersions is the {\it extended
dispersion}.

Before proceeding to show the average properties of the cluster
ensemble we examine in some detail the structural and thermal 
properties of three typical clusters, selected with the intention to
approximately cover the range of concentrations apparent in the
ensemble. Fig.~\ref{fig:select3} displays radial profiles,   
namely the density, radial bulk velocity, velocity dispersion and entropy 
profiles, of three arbitrarily chosen clusters of comparable masses
($\sim 10^{15}\hMsol$). The uppermost 
panels show the density profiles for the DM and the gas scaled by the
corresponding cosmic mean density . The concentration parameter
$c=R_{\rm vir}/R_s$ specified in the upper right corners is based on a
NFW-fit \citep{Navarro97} to the DM component and the associated
profile is displayed with the dotted lines. The gas velocity
dispersion and entropy is decomposed into the thermal and kinetic
components. In addition the extended dispersion and entropy of the 
gas is shown.

The important features to notice in Fig.~\ref{fig:select3}
are: (a) The excellent NFW fit to the DM density profile of the
high concentration cluster, while the gas density exhibits a core
structure; (b) The lack of significant radial streaming motions of the
high concentration cluster implies that it is a relaxed cluster while
the others are still dynamically evolving; (c) The DM velocity
dispersion profile of all clusters show a 'temperature 
inversion', namely the velocity dispersion increases with radius
within $\approx 0.1R_{\rm vir}$. The gas thermal and extended velocity
dispersions at the very centre (below $\approx 0.05R_{\rm vir}$) show 
the opposite behaviour, at these radii a negative temperature gradient
of the gas is apparent in all the clusters. Cluster $c=6.9$ displays
an almost isothermal core with the hint for a maximal temperature at
the same radius where the DM dispersion peaks. (d) The DM entropy
profile is close to a power law independent of the dynamical state of
the cluster (compare to \citealt{Ascasibar04}). Within $\approx
0.1R_{\rm vir}$ the thermal and extended gas  entropies show a core 
structure. Outside that region the extended gas entropy almost
perfectly  coincides with the DM entropy.   

\begin{figure}
\epsfig{file=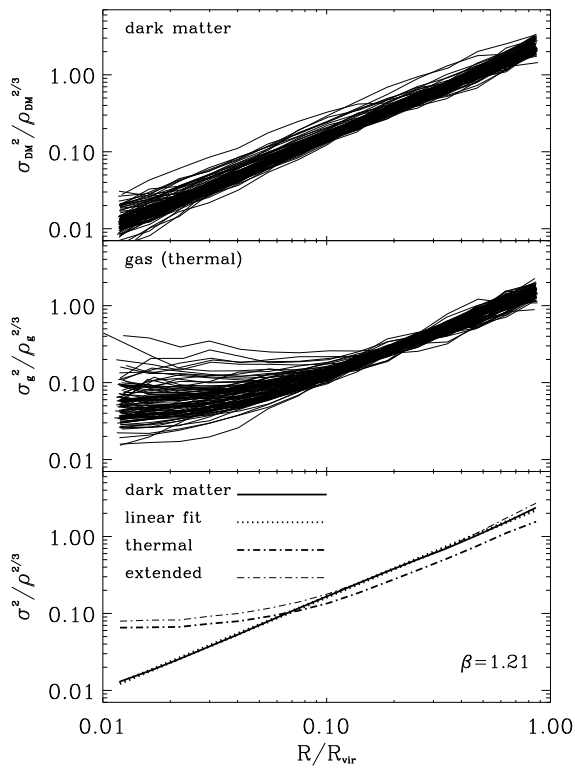,width=0.95\hsize}
\caption{\label{fig:scam99}
The upper two panel display the DM and the gas entropy profiles of the
clusters within the ensemble. To reduce the scatter caused by the
varying masses of the clusters the densities have been scaled by their
mean values within the virial radius and the velocities have been
scaled by the virial circular velocities 
$(\G M_{\rm vir}/R_{\rm vir})^{1/2}$. The lowest panel shows the mean
DM and gas entropies by the thick solid and dashed-dotted lines. The
thin dashed-dotted line is the mean profile of the extended entropy of
the gas. The dotted line on top of the solid line represents the best
fitting power law for the DM component with a logarithmic slope of
$\beta = 1.21$. The corresponding exponent for the phase space density
is $\alpha = -1.82$.
}
\end{figure}
Fig. ~\ref{fig:scam99} presents a direct comparison of the DM
and the thermal and extended gas entropy profiles as defined in Eq.~\ref{equ:Ki} 
for the 96 clusters in the sample. The DM entropy is displayed in the
upper panel and the gas thermal entropy is  given in the middle
panel. The bottom panel exhibits the mean of the DM and the thermal
and extended gas entropy profiles. The mean profiles reproduce the
trends followed by the individual clusters of
Fig.~\ref{fig:select3}. The DM profile follows a power law of the form
$K_{\rm DM}\propto r^\beta$, where $\beta=1.21$. This logarithmic
slope   corresponds to a slope for the phase space density  profile of
$\alpha = -1.82$. The mean gas thermal and extended profiles keep
almost a constant displacement (in the log-log plane). The gas
exhibits an entropy core of a radius $R_{\rm core}\approx 0.1 R_{\rm
vir}$. Outside that core the gas entropy trails the DM power law. In
particular, the extended entropy coincides with  the DM entropy.
The lower panel of Fig. ~\ref{fig:scam99} presents the main new result
of the paper, namely that the gas extended entropy outside the core
coincide with the DM entropy not only in slope but also in amplitude.  

\begin{figure}
\epsfig{file=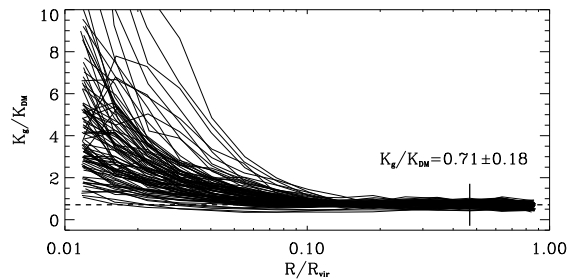,width=0.95\hsize}
\caption{\label{fig:horizo99}
Ratio of gas and DM entropies. The mean value of
$K_{\rm g}/K_{DM}$ at the radius indicated by short vertical
line is displayed by the dashed horizontal line.} 
\end{figure}
The thermal entropy of the gas is more accessible to astronomical
observations than the extended one. So, the ratio of the thermal to
the DM entropy profile of all the clusters is presented
Fig.~\ref{fig:horizo99}. Outside the core that ratio is virtually
constant and exhibits a small scatter,  
$K_{\rm g}/K_{\rm DM}=0.71\pm 0.18$.  

\section{Gas Core Structure}
\label{sec:core}
\begin{figure}
\epsfig{file=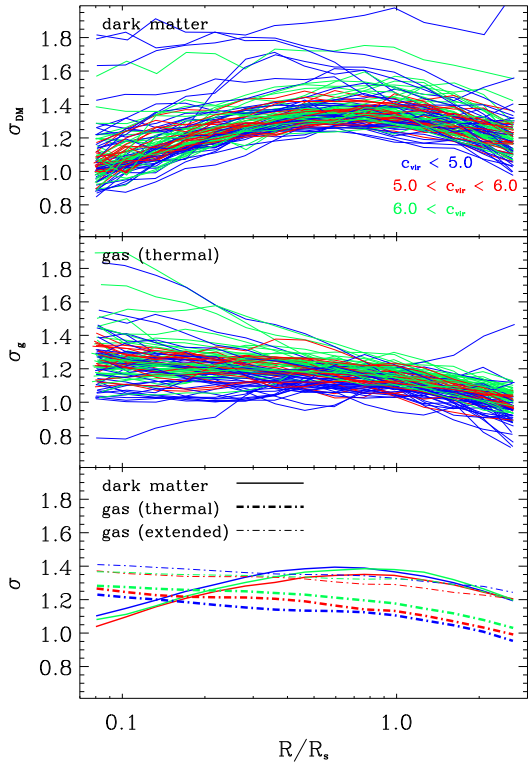,width=0.95\hsize}
\caption{\label{fig:stemp99}
The upper two panel show the temperature   profiles for the DM and
the gas, respectively, of the clusters sample. The profiles are
plotted against the radius scaled by the NFW \rs. The different line
styles of the profiles are chosen according to the concentration
parameter, where clusters are grouped into $c<5$, $5 < c < 6$ and $6 <
c$. The bottom panel shows the mean profiles, where the mean is taken
over the groups with the above $c$ range. Both the thermal and
extended temperature profiles are presented. The DM 'temperature'
(namely velocity dispersion) shows a clear temperature inversion
within $\approx 0.8 R_{\rm s}$.  The gas profiles is slightly
monotonically decreasing within that inner region.  
}
\end{figure}
\begin{figure}
\epsfig{file=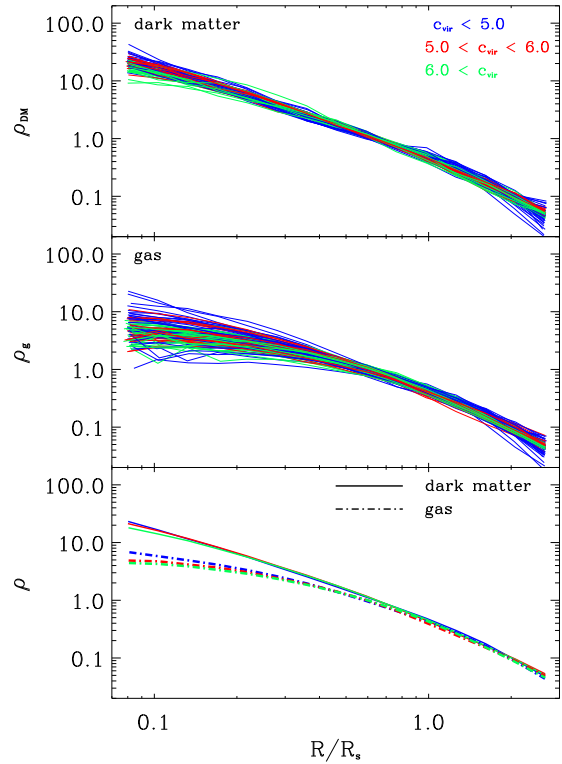,width=0.95\hsize}
\caption{\label{fig:sden99}
The upper two panel show the scaled density  profiles for the DM and 
the gas, respectively, of the clusters sample. The profiles are
plotted against the radius scaled by  \rs. The line styles follow
the one of Fig.~\ref{fig:stemp99}. The bottom panel displays the mean 
profiles, where the mean is taken for each concentration bin 
separately.   
}
\end{figure}
The present results provide further support for the findings by
\cite{Voit05} that the emergence of the entropy core in adiabatic
simulations is a real physical effect and is not a numerical
artifact. Another indication for the reality of the entropy core comes
from the comparison of SPH with Eulerian grid
codes. \cite{Ascasibar03} found good agreement between the same  
cluster simulated by both techniques. In particular 
an entropy core emerged in both realisations. However, as noted by \cite{Voit05b} the  
core entropy levels can differ by almost a factor of two. Also,
\cite{Lin06} point out that the resolution of the SPH 
simulations plays an important role. Yet, the resolution for
the ensemble presented here is sufficient to exclude the pure artificial
nature of the entropy cores. This statement is based in particular on
the intermediate and low concentration systems. Here the radius of the
core region is $\gtrsim200\hkpc$ (assuming a virial radius of $1\hMpc$
and a concentration of 5) which can not be a caused by a lag of spatial
resolution. It follows that the mere existence of
entropy cores does not imply non-adiabatic physics such as feedback
and preheating. 

\begin{figure}
\epsfig{file=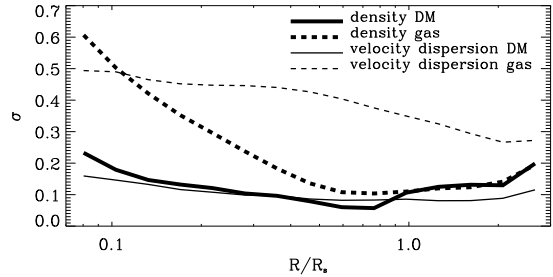,width=0.95\hsize}
\caption{\label{fig:sca}
Radial dependence of the scatter for the density and velocity profiles
of the gas and the DM. Displayed are the rms values of
$(x-\bar{x})/\bar{x}$, where $x$ indicates the values associated with
the individual profiles and $\bar{x}$ is their mean value within the
bin. The thick solid and dashed lines show the scatter in the gas and
DM densities (Fig.~\ref{fig:sden99}). The thin solid and dashed lines
depict the scatter in the gas and DM velocity dispersions
(Fig.~\ref{fig:stemp99}).
}
\end{figure}

To improve our understanding of
that phenomenon we undertake a more detailed investigation of entropy
core.  As we are interested in the self-similar properties of the
clusters all further plots of radial profiles will be scaled by the
NFW scale radius \rs, rather than by \rvir. At the same time the
density profiles are scaled by their mean densities within \rs\ and
the velocities are scaled by the circular velocity at \rs , $v_{c,s} =
(\G M_s/R_s)^{1/2}$, where $M_s$ is the total mass within \rs. 
That kind of scaling achieves a remarkable agreement among the dark
matter density profiles, implying that the scatter of the gas density
profiles, in particular at the centre, is a real physical effect
(Figs.~\ref{fig:stemp99},~\ref{fig:sden99} and ~\ref{fig:sca}). 
To quantify the scatter in the density and velocity dispersion
profiles we compute the rms value of $(x-\bar{x})/\bar{x}$, where
$x$ indicates the values associated with the individual profiles and
$\bar{x}$ is their mean value within a particular bin. As shown in
Fig.~\ref{fig:sca} the scatter of the DM density lies between $0.2$
and $0.1$ and a minimum is located near $R_s$. The central scatter
$\rho_{gas}$ is $0.6$ but it declines with radius and converges to
the DM scatter at $\sim0.8 R_s$. The scatter in $\sigma_{DM}$ is very
similar to that of the DM density profile, ranging in between to $\sim
0.2$ to $\sim 0.1$. The scatter in the thermal velocities of the gas
ranges from $0.5$ at the centre to $0.3$. It follows that there is a
much larger variation among the different clusters in the gas density
and temperature profiles than in the DM component. 
To explore the dependence of the core structure on
the dynamical state of the cluster the ensemble has been subdivided
into three concentration bins. The three bins, namely $c < 5$, $5 < c
< 6$ and $6 < c$ comprise 39, 19 and 38 clusters, respectively. The
intermediate bin should be considered as a kind of a buffer zone for
the statistically more robust bins on the edges. The clusters with
concentrations  below 5 are assumed to be on average dynamically young
systems whereas the clusters with concentrations above 6 are expected
to be the most relaxed systems within the ensemble (see
e.g. \citealt{Wechsler02}).  
Under the assumption that the ICM is in hydrostatic equilibrium 
the gas entropy excess within $R_{\rm core}$ implies that the gas is 
hotter and/or the gas density profile is flatter than the DM. 
This is confirmed by the simulation. The radial structure of
the DM and gas temperature is given in Fig.~\ref{fig:stemp99}. The
figure nicely illustrates the temperature inversion of the DM
temperature profile and, in contrast, the gentle increase of the gas
temperature towards the centre. Fig.~\ref{fig:sden99} presents the
DM and gas density profiles and it clearly shows the flattening of the
inner gas density profile.  The gas entropy core is established by the
flattening of the NFW density cusp and the absence of the temperature
inversion.

\begin{figure}
\epsfig{file=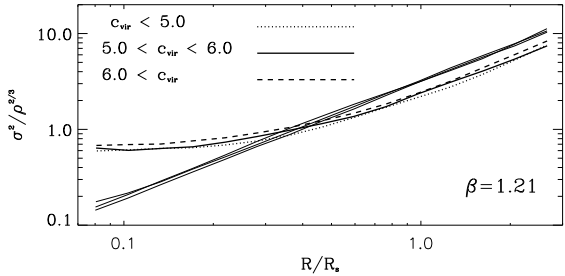,width=0.95\hsize}
\caption{\label{fig:cora99sgau13}
The mean DM and gas  thermal entropy profiles is presented for the
different concentration parameter grouping. The DM entropy file
follows the $r \propto r^\beta$ power law, independent of the value of
$c$. 
}
\end{figure}
Next we focus on the impact of the concentration parameter on the
entropy profiles. Fig. ~\ref{fig:cora99sgau13} shows the 
mean DM and gas thermal entropy profiles and their dependence on the
value of $c$. The DM is found to be independent of $c$ and the gas
shows a very marginal dependence. A much stronger variation of the
core entropy level and the core sizes with concentrations is seen if
$R_{vir}$-scaled  quantities are used (not shown here). In that
context it is worth noting, 
that the mean density profiles for the different concentrations do
match very well apart from the inner 0.3\rs\, where the lowest
concentration profiles are slightly enhanced
(Fig.~\ref{fig:sden99}). In contrast the mean temperature profiles of
the gas (Fig.~\ref{fig:stemp99}) show a slight dependence on
concentrations, namely the gas temperature rises with
concentration. This trend is not reflected in the behaviour of the DM
velocity dispersion profiles and therefore it may be real and not
caused by the $v_{c,s}$ scaling. Therefore, higher concentrated
clusters have slightly higher gas temperatures. Now, rising
temperatures at equal densities result in higher entropies for higher
concentrated clusters as seen in Fig.~\ref{fig:cora99sgau13}. However,
it can be questioned whether the scaling of the velocities by the
circular velocity at \rs\ ($v_{c,s}$) is the physically appropriate
way of gauging the temperature profiles of different sized clusters.

\begin{figure}
\epsfig{file=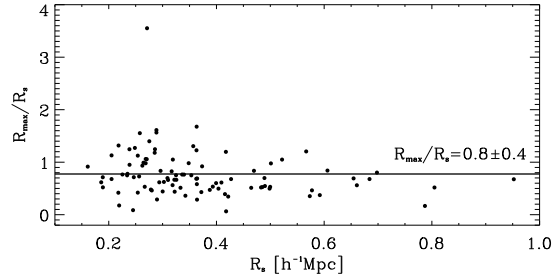,width=0.95\hsize}
\caption{\label{fig:rsrs99tgau13}
Distribution of $R_{\rm max} /  R_{\rm s}$ {\it vs} \rs, where $R_{\rm
max}$ is the radius at of the maximum of the DM 'temperature'. The
distribution is characterised by $R_{\rm max} /  R_{\rm s} =  0.8 \pm
0.4$. 
}
\end{figure}
The core radius seems to coincide with the radius at which the DM
temperature reaches its maximal value, defined as $R_{\rm max}$. Thus
it seems that the 'thermal', namely velocity dispersion  of the DM, 
properties of the DM either set, or otherwise are related to, the
extent of the entropy core. Fig. ~\ref{fig:rsrs99tgau13} presents a
scatter plot of the ratio $R_{\rm max} /  R_{\rm s}$ versus
\rs. There is no significant dependence of $R_{\rm max} /  R_{\rm s}$
on \rs\ or on $c$. Associating $R_{\rm max}$ with the core radius we
conclude that  $R_{\rm core} / R_{\rm s} = 0.8 \pm 0.4$.  
\section{Discussion}
\label{sec:discu}
The main findings of the paper are: The DM phase space density power
law of \cite{Taylor01} is reproduced by the simulated clusters
extracted from an adiabatic gas dynamical simulation. The 
gas entropy profile is in accord with previous simulations
\citep{Voit05b}. In particular the  entropy profile has an inner flat 
core and an outer profile of $K_{\rm g}\propto r^{1.21}$ is
recovered. This stands in at least qualitative agreement with X-ray
observations. The entropy core coincides with the region of the
temperature inversion of the DM and its radius  is  characterised by
$R_{\rm core} /  R_{\rm s} = 0.8 \pm 0.4$.  However, entropy loss due
to radiative processes, which is not included in our simulations,
alters the thermodynamics of the central ICM, see
e.g. \cite{Kravtsov05}. Therefore, it is no contradiction to our
findings that \cite{Piffaretti05} don't detect entropy cores within 13
nearby cooling flow clusters. The most important result of the paper
is that outside the entropy core the gas and DM entropies follow one
another very closely. Upon appropriate scaling of the gas and DM
densities by their cosmological mean values we find that outside the
entropy core a constant ration of $K_{\rm g}/K_{\rm DM}=0.71\pm0.18$
prevails. By extending the definition of the gas entropy to include
also the kinetic energy of the turbulent motion the above ratio
becomes unity, and it follows that in the outer part of the cluster
the gas and DM entropies coincide and are virtually the same.   

The constant ratio of the gas to the DM entropies outside the entropy
core has interesting observational ramifications. To the extent that
non-adiabatic physics, in the sense described above, does not affect
the gas structure outside the entropy core observations of the
ICM can shed light on the DM in clusters. Observational determination
of the entropy of the ICM can be easily translated into the entropy of
the DM.  Given the power law nature of the DM entropy, the
observationally determined outer DM entropy can be extrapolated
inwards. This, together with independent mapping of the mass
distribution can yield interesting insight into the distribution and
dynamics of DM in clusters of galaxies. 
In this context it is worth mentioning that \cite{Ikebe04} described 
a formalism how the DM ``temperature'' can be derived from X-ray
observations and applied this new technique to {\it XMM-Newton} data
of A1795. 
The analysis of \S~\ref{sec:core} shows that the entropy core radius,
measured in \rs\ units, is independent of the cluster NFW
parameters. This seems to imply that the emergence of the entropy core
is related to processes occurring within \rs. The common understanding
of halo formation is that the structure within \rs\ is determined
exclusively in the violent merger phases the cluster undergoes
\citep{Wechsler02, RomanoDiaz06}. The quiescent accretion
phases that follows the violent mergers do not alter the inner
structure but leads to a continuous growth of virial radius. This
suggests that the entropy core emerges as a result of processes
occurring during the violent phases, in which the cluster is not in
NFW equilibrium. Another alternative is that the gas component, unlike
the DM, undergoes some secular evolution of the inner region in the
quiescent phases. It might also be the case that both processes play
an important role. In any case it is clear that only shock waves
can form the entropy core and the understanding of 
the shock pattern is critical here.

N-body simulations show that clusters have experienced merging
events in their past. The merging events of comparable sized
substructures are violent enough to send bow shocks into the
intra-substructure gas which subsequently  propagate to the outskirts
of the  clusters \citep{Nagai03}. This shocks raise the
entropy of the gas. \cite{Pfrommer06} develop a formalism for the
identification and accurate estimation of the strength of structure
formation shocks during cosmological SPH simulations. They find that
most of the energy is dissipated in weak internal shocks with Mach
numbers $\mathcal{M}=2$ which are predominantly central flow shocks or
merger shock waves traversing halo centres. The study by
\cite{Navarro95} utilises cosmological simulations to compare the
evolution of DM and gas of clusters between $z=1$ and $z=0$. They find
that the DM evolves at roughly constant phase space density, whereas
shocks increase the temperature of the gas with relatively little
change in density. \cite{Ascasibar04} investigating pure N-body
simulations  report the remarkable stability of the phase space density
profile of the DM independent of the dynamical state.
\cite{Pearce94} explore simple head-on mergers between model galaxy
clusters containing a mixture of gas and dark matter, testing the
coupling of the gas to the underlying collisionless material. After
the first passage of the dark matter cores the gas forms a
pancake-like structure about the centre of mass. While separating for
a second time dark matter cores leave their gas behind. The central
gas is no longer confined to a deep potential well so it expands.
This process has been nicely confirmed by the recent observations 
of \cite{Clowe06}. They find an obvious displacement of the gas measured
by X-ray observations compared to the potential revealed by weak
lensing maps of the merging cluster system 1E 0657-558.  
At the same time gravitational attraction due to
the central gas exerts a drag force on the separating DM cluster
cores, causing an energy transfer from the DM to the gas (see also 
\citealt{Navarro93}). This is in agreement with recent investigations 
by \cite{Lin06} reporting a $\sim10$ per cent increase in the
concentration of the cluster's DM component when adiabatic gas physics
is included as compared to pure DM simulations.  The core structure of
the gas entropy, in contrast to the power-law profile of the DM, may
therefore be established during merging events in consequence of the
impenetrability of the gas as opposed to the collisionless nature of
the DM.   

A different point of view  might be provided by the following
speculations. The DM 'temperature' profile is characterised by a
temperature inversion. Now, a 'normal' thermodynamic system with a
mechanism for heat transport and no energy sinks or sources cannot
sustain such a structure. Heat would flow inwards, leading to an
increase of the entropy where the temperature inversion prevails. The
increase in entropy amounts to an increase in temperature, and thereby
decrease in density so as to keep the hydrostatic equilibrium. Now,
let us conjecture that the collapse of the gas and DM results in a 
configuration where the gas follows the DM distribution. (This  is
supported by the self-similar solutions of Bertschinger 1985 and
Chuzhoy \& Nusser 2000). If the gas had a mechanism of transporting
energy it would behave in the manner described above, namely the
entropy   will increase  where the initial temperature inversion
initially existed. Inspection of Fig.~\ref{fig:select3} proves that
this might have actually happened. The gas entropy departs from the DM
and an entropy core replaces the temperature inversion. This is
clearly manifested by the most relaxed cluster of
Fig.~\ref{fig:select3}, and  is seen in   other relaxed clusters
(not shown here). So, what can provide the heat transport mechanism?
Fig.~\ref{fig:select3} shows that outside the entropy core about
$(25-30)$ per cent of the non-gravitational energy is in the form of kinetic
energy. Only a small fraction of this energy is in large scale bulk
motion and therefore most of it is in small scale motion, namely
turbulence. The dissipation of these motions by weak shocks is
expected to be efficient at the centre and this would lead to the
increase in the gas entropy. In the language of one dimensional
spherically symmetric hydrodynamics one would attribute this
phenomenon to turbulent viscosity and heat transport. This transport 
mechanism erases the temperature inversion and produces the entropy
core. 

The application of   the concept of maximum entropy  as a mean of
calculating the equilibrium configuration   of   self gravitating
systems has proven somewhat  futile
(e.g.~\citealt{LyndenBell67,LyndenBell68,White87,LimaNeto99}). In  
particular,   these attempts   have not predicted the 
NFW density profile that  faithfully reproduces the structure of DM
halos. The problem that lies at the root of the shortcoming of the
thermodynamic approach is the long range nature of the gravitational
force and the resulting negative specific energy (see
e.g.~\citealt{LyndenBell68}). In spite of all that our results show
that the concept of the  DM entropy is closely related to
the structure of DM halos. Moreover, the DM entropy is very closely
related to the gas entropy. Beyond the entropy core of the gas the
entropies of the two components are tracing one 
another. In fact,  the extended gas entropy fully coincides with the
DM entropy in that regime. It follows that the observed power law
behaviour of the gas entropy provides an indirect support for the NFW
structure of the DM. Now, this opens the door for a fundamental
question. The gas entropy results from the conversion of gravitational
to kinetic energy and by the dissipation of the kinetic energy into
internal one by means of shock waves. The power law regime seems to be
dominated by the accretion shock that is predicted by self-similar
solutions \citep{Bertschinger85,Chuzhoy00}. In the DM sector  the 
gravitational energy is converted to random motions  by means of phase
mixing and the time variation of the  gravitational potential. The latter
process happens by violent relaxation, whereby the gravitational
potential changes rapidly, or by means of adiabatic changes as
envisaged in the secondary infall model \citep{Gunn77, Fillmore84,
Bertschinger85, Hoffman85}. The physical processes that shape the
entropy profile in the gas and the DM appear to be very different, yet
the resulting structure seems to be identical. This phenomenological
finding  needs to be further investigated. 

Additional studies of the adiabatic formation of clusters of galaxies
are needed to deepen our understanding of the role of entropy in the
formation and evolution of clusters of galaxies. In particular, high
resolution gas dynamical simulations need to be analysed with a
special emphasise on the role of strong, large scale, shocks and of
weak, small scale, ones. The relevance of these to the emergence of
the entropy core needs to be further clarified. Finally, the
simulations should be expanded so as to include preheating, radiative
cooling and feedback processes.  

\section*{Acknowledgements}
Useful discussions with Y. Ascasibar, M. Hoeft, A. Klypin,
A. Kravtsov and W. Mathews are highly appreciated. We are
grateful to Hans B\"ohringer for the very constructive referee report.
This research has been supported by NSF grant AST 00-98351 and NASA
grant NAG5-13275 (to AF) and by  ISF-143/02 and the Sheinborn
Foundation (to YH). AF acknowledges  the hospitality of the Hebrew
University. YH acknowledges DFG for a Mercator Gastprofessur at
Potsdam University. GY thanks  M.E.C (Spain) for financial support
under projects  AYA-2003-07468 and  BFM2003-01266. GY and SG thanks
the Acciones Integradas Hispano-Alemanas for support. We would like to
thank the Barcelona Supercomputer Center for allowing us to perform
this simulation during the testing period  of the MareNostrum
supercomputer. The simulation has been analysed at NIC Juelich. 

\end{document}